\documentclass[final,5p,times,twocolumn,numeric]{elsarticle}

\usepackage{amssymb} %% The amsthm package provides extended theorem environments
\usepackage{xcolor} % coloured text
\usepackage{amsthm}
\usepackage{mathtools}
\usepackage{abraces}
\usepackage{moreverb,url}
\usepackage{amsmath,bm}	
\usepackage[ngerman, english]{babel}

\usepackage{algpseudocode}
\usepackage[colorlinks,bookmarksopen,bookmarksnumbered,allcolors =blue]{hyperref}
\usepackage{tabularx}
\usepackage{booktabs}
\usepackage{xstring}
\usepackage{siunitx}
\usepackage{caption, subcaption}
\usepackage{pgfplots}
\usepackage{pgfplotstable}
\pgfplotsset{compat = newest}
\usepackage{filecontents}
\usepackage{tikz}
\usepackage{tikzscale}
\usepackage{tikz-3dplot}

\newcommand{\tensor}[1]{\IfSubStr{ABCDEFGHIJKLMNOPQRSTUVWXYZabcdefghijklmnopqrstuvwxyz}{#1}
        {\mathbf{#1}}
        {\bm{#1}}}

\journal{Ultrasonics}

\begin{document}

\begin{frontmatter}



\title{Quantitative Comparison of the Total Focusing Method, Reverse Time Migration, and Full Waveform Inversion for Ultrasonic Imaging}

\author[b]{Tim Bürchner$^*$}
\author[a]{Simon Schmid$^*$}
\author[a]{Lukas Bergbreiter$^*$}
\author[c]{Ernst Rank$^\text{b,}$}
\author[d]{Stefan Kollmannsberger}
\author[a]{Christian U. Grosse}

\affiliation[a]{
organization={Technical University of Munich, TUM School of Engineering and Design, Department of Materials Engineering, Chair of Non-destructive Testing},
addressline={Franz-Langinger-Str. 10}, 
city={Munich},
postcode={81245}, 
country={Germany}
}

\affiliation[b]{
organization={Technical University of Munich, TUM School of Engineering and Design, Chair of Computational Modeling and Simulation},
addressline={Arcisstrasse 21}, 
city={Munich},
postcode={80333}, 
country={Germany}
}

\affiliation[c]{
organization={Technical University of Munich, Institute for Advanced Study},
addressline={Lichtenbergstrasse 2a}, 
city={Garching},
postcode={85748}, 
country={Germany}
}

\affiliation[d]{
organization={Bauhaus-Universit\"at Weimar, Chair of Data Science in Civil Engineering},
addressline={Coudraystraße 13}, 
city={Weimar},
postcode={99423}, 
country={Germany}
}
% \affiliation[d]{
% organization={equal contribution}
% }

\begin{abstract}
Phased array ultrasound is a widely used technique in non-destructive testing. 
Using piezoelectric elements as both sources and receivers provides a significant gain in information and enables more accurate defect detection. 
When all source-receiver combinations are used, the process is called full matrix capture. 
The total focusing method~(TFM), which exploits such datasets, relies on a delay and sum algorithm to sum up the signals on a pixel grid. 
However, TFM only uses the first arriving p-waves, making it challenging to size complex-shaped defects. 
By contrast, more advanced methods such as reverse time migration~(RTM) and full waveform inversion~(FWI) use full waveforms to reconstruct defects. 
Both methods compare measured signals with ultrasound simulations. 
While RTM identifies defects by convolving forward and backward wavefields once, FWI iteratively updates material models to reconstruct the actual distribution of material properties. 
This study compares TFM, RTM, and FWI for six specimens featuring circular defects or Y-shaped notches. 
The reconstructed results are first evaluated qualitatively using different thresholds and then quantitatively using metrics such as AUPRC, AUROC, and F1-score. 
The results show that FWI performs best in most cases, both qualitatively and quantitatively.
\end{abstract}


\end{frontmatter}


\section{Introduction}
\label{intro}
\def\thefootnote{*}\footnotetext{These authors contributed equally to this work}

Ultrasonic testing with single element or phased array transducers is a well-established non-destructive testing (NDT) method. Phased array ultrasonic testing provides advantages toward ultrasonic testing with a single probe such as shorter inspection times, a simple testing setup, and a better defect characterization. Accurately estimating the defect size is usually essential for assessing the criticality. Fracture mechanics simulations can exploit this information to predict the effect of the identified defects~\cite{BSI1}  under certain loads and other boundary conditions on the remaining lifetime of the component. However, due to physical and algorithmic constraints, estimating the size of defects accurately with ultrasound remains challenging, making it an area of ongoing research. Different techniques rely on different ultrasound signal characteristics, such as the amplitude~\cite{Kleinert, Ono}, the time-of-flight~\cite{Ataby}, or the full waveform~\cite{Fichtner}. Imaging approaches typically employ single time domain signals forming B- or S-Scans, delay and sum algorithms like the total focusing method (TFM)~\cite{Holmes} or simulations like reverse time migration (RTM)~\cite{Baysal1983} and full waveform inversion (FWI)~\cite{Tarantola1984}. RTM and FWI are well known in geophysical applications, but rarely used in NDT applications. The paper at hand comprises qualitative and quantitative comparisons of TFM, RTM, and FWI. The study uses full matrix capture (FMC) for data acquisition and successfully applies FWI to the acquired datasets.

In the past, extensive research has been conducted to improve TFM by considering factors such as anisotropy~\cite{Menard, Grager}, heterogeneity~\cite{Singh, Luo}, or focusing on near-surface defects~\cite{Bergbreiter}.
Different sound paths and mode conversions can be explicitly implemented in TFM, allowing multi-mode imaging to improve defect detection and characterization in different measurement setups~\cite{ZHANG2010}. 
In summary, TFM images based on single wave modes are challenging to interpret because signals from different sound paths and mode conversions (backwall skip, multiple reflections, ...) create complex indications such as shadow reflectors in the reconstruction image, and thus require additional treatment~\cite{Schmid2023}.
Direct TFM, as applied in the presented work, is typically used in industry and considers only one wave mode (longitudinal wave or shear wave). 
Mode conversion and other wave types have to be incorporated explicitly.
By contrast, simulation-based algorithms implicitly account for all wave modes and conversions contained in the underlying wave equation, improving the imaging significantly. 
For more information on FMC and TFM, please refer to Section~\ref{FMC} and Section~\ref{TFM}. 

RTM or elastic RTM are simulation-based imaging approaches initially developed in geophysics~\cite{Chang}. Elastic RTM considers all wave types included in elastic wave simulations for the reconstruction. Studies~\cite{Grohmann, Rao2} show that elastic RTM provides a better reconstruction of the underlying structure than TFM or the synthetic aperture focusing technique (SAFT). SAFT is a method closely related to TFM that only uses the diagonal entries of the information matrix (see Section~\ref{FMC}). For an overview of RTM theory see Section~\ref{RTM}. Grohmann et al.~\cite{Grohmann} explore 2D elastic RTM in a study using synthetic data where they identify the back and side walls of a concrete specimen with staircase features. 
The simulations are performed using the finite difference method and a phased array model of p-wave transducers. 
SAFT is also used for reconstruction and compared to RTM. 
The RTM images show more detail than the SAFT images for the defects investigated, such as vertical boundaries and lower edges of circular voids. The authors conclude that 2D elastic simulations using p- and sv-modes, along with surface waves, are suitable to accurately capture complicated object structures.
Further studies on 2D elastic RTM for ultrasound measurement data are presented in~\cite{Buttner, Zhang, Mizota2022, Liu}. Rao et al.~\cite{Rao2} compare TFM images to 2D elastic RTM images for two complex shaped notches based on phased array measurements. The results of their work show that elastic RTM was able to capture the shape of the notches more accurately than TFM. 
This is likely due to the ability of elastic RTM to utilize more information in the reconstructions, particularly sv-wave mode information.

FWI is also a simulation-based method. But unlike RTM, it iteratively updates material models, such as wave speeds and densities, within an optimization framework. 
In other words, FWI fits the data from simulated signals to measured signals. 
This process results in an image that mimics the mechanical behavior of the sample of interest. 
FWI is widely used in industry for geophysical applications (e.g., oil and gas exploration)~\cite{Fichtner}.
For more information on FWI, see Subsection~\ref{FWI}. 
In ultrasonic testing, FWI leverages all information contained in the corresponding wave simulations, including multiple reflections from boundaries and mode conversions. 
In guided wave tomography, Rao et al.~\cite{Rao4} apply FWI to determine the remaining thickness of aluminum plates. 
They solve the wave equation in the frequency domain and use dispersion curves to map the resulting wave speeds obtained in the inversion to plate thicknesses. 
Chen et al.~\cite{Chen} apply 2D elastic FWI to locate rebars in concrete slabs using shear wave transducers to reconstruct the shear wave speed $v_\textrm{s}$ along with the density. 
The results show that FWI identifies the rebar sizes more accurately than SAFT and ground-penetrating radar, which fail to estimate the rebar sizes correctly. 
Seidl et al.~\cite{Seidl} use FWI with synthetic data to estimate the location of rebars in a 3D concrete block, using acoustic wave simulations and first arriving p-waves. 
Other studies using FWI in ultrasonics are given in~\cite{Luan, Rossato, Schmid2024, Krischer}.
In contrast to most of the literature, the paper at hand investigates more complex shaped defects, using high-frequency transducers~($\SI{2.25}{\mega \hertz}$). 
To the best of the authors' knowledge, this is the first study to apply classical FWI to phased array FMC datasets in ultrasonic defect detection. 
The defects investigated are voids that mimic blowholes or cracks, and are partially inspired by those used in Rao et al. ~\cite{Rao1}. 
Three different reconstruction methods, namely TFM, RTM, and FWI, are studied and compared qualitatively and quantitatively using different segmentation metrics. 
The quantitative comparison of reconstruction method for ultrasonic testing represents also a novelty of this paper.

The remainder of the paper is structured as follows. 
Section~\ref{METH} provides the relevant background regarding the employed methods, namely FMC, TFM, RTM, and FWI. 
Subsequently, Section~\ref{SETUP} outlines the experimental, simulation, and optimization setup. 
Section~\ref{RESULTS} presents and discusses the results, including the reconstruction outcomes of all three methods for two defect variants: circular holes and Y-shaped notches. 
The results are compared qualitatively and quantitatively, followed by concluding remarks in Section~\ref{CON}.

\section{Methodology}
\label{METH}

\subsection{Full Matrix Capture (FMC)}
\label{FMC}

\begin{figure}[htbp!]
  \begin{center}
    \includegraphics[width=0.45\textwidth]{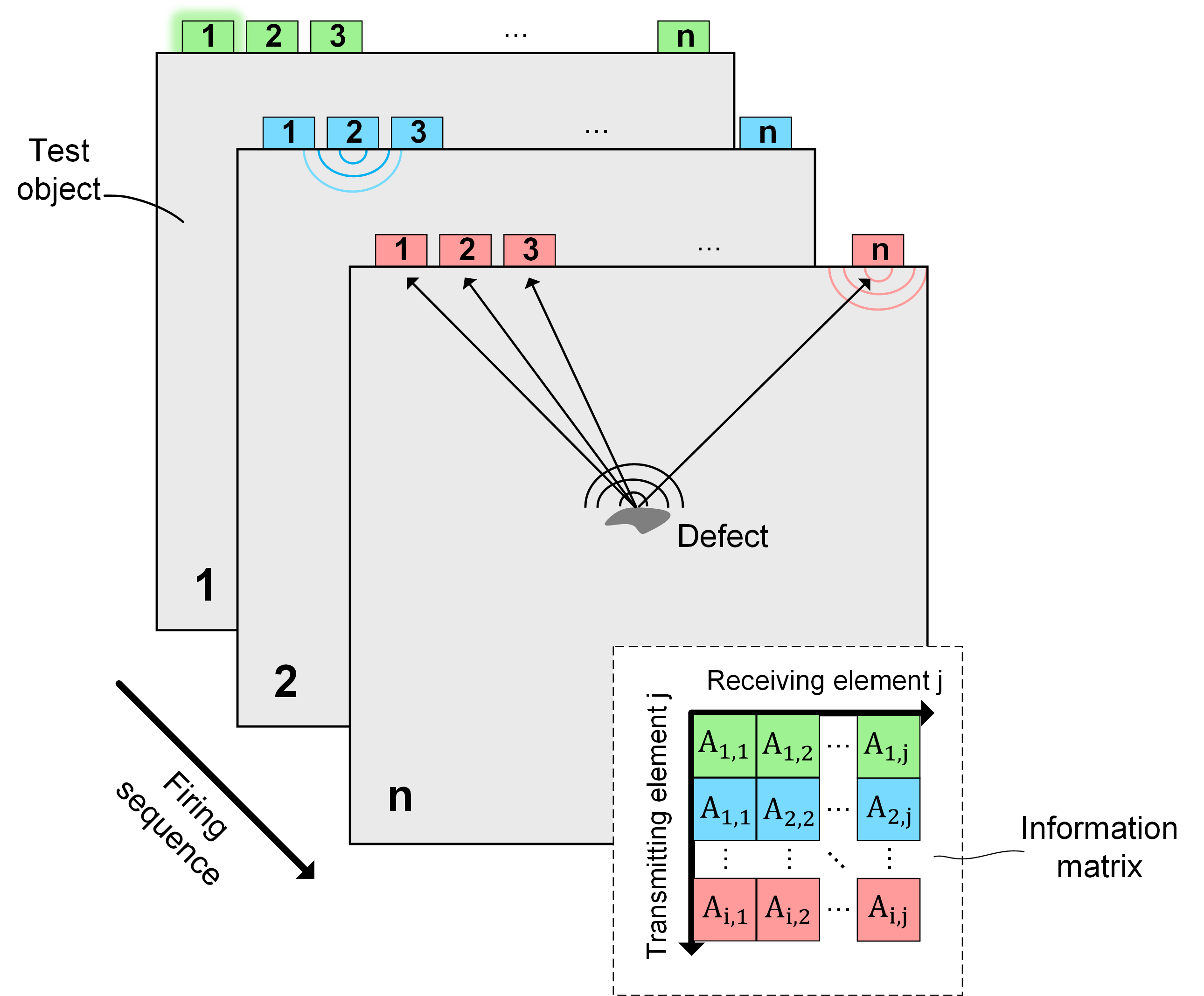}
  \end{center}
  \caption{FMC acquisition (adapted from Schmid et al. ~\cite{Schmid2023})}
  \label{fig:FMC}
\end{figure}

Phased array probes are widely used in industry for conventional ultrasonic testing. FMC is commonly applied to acquire data to fully exploit the potential of ultrasonic phased array transducers. FMC captures the time-domain signals (A-Scans) for all possible transmitter-receiver combinations of phased array probes with $n$ elements. The resulting data is arranged in a $n \times n$ information matrix $I_{i,j}$, where the index $i,j$ denotes the A-scan at receiver $j$ when the transmitter $i$ is excited. In the post-processing phase, various algorithms, including TFM, RTM, and FWI, can be applied to the FMC data~\cite{Holmes}. The FMC procedure is sketched in Figure~\ref{fig:FMC}.

\subsection{Total Focusing Method (TFM)}
\label{TFM}
The TFM is the most widely used post-processing algorithm for FMC data in ultrasonic testing. The grid-based TFM achieves higher resolution than conventional imaging by focusing synthetically on every pixel of the defined region of interest~(ROI). TFM calculates the traveling times of every transmitter-receiver combination in the information matrix $I$ to all pixels in the grid. Subsequently, TFM delays the amplitudes of the single A-Scans by the travel time and sums them up. This delay and sum leads to constructive interference at the positions where a reflector is present and to destructive interference everywhere else. The summation process suppresses the noise of the single A-scans, inherently resulting in a higher signal-to-noise ratio (SNR) than in conventional ultrasonic scans. This paper investigates complex TFM, which applies the Hilbert transformation to the information matrix. Only 2D TFM is considered for simplicity, which is an adequate approximation in ultrasonic testing~\cite{Holmes}.

The TFM pixel intensity is:
\begin{equation}
I(x,z) =\left|\sum_{i}^{N} \sum_{j}^{N}X_{ij}\left ( t = \frac{s_{1}+s_{2}}{c} \right )\right|
\label{eq:tfm}
\end{equation}
where $x$ is the pixel coordinate in index direction, $z$ the pixel depth, $N$ the phased array probe element count, $i$ the transmitter index, and $j$ the receiver index. The Hilbert transformed information matrix entry is denoted by $X_{ij}$, where $s_1$ is the sound path from transmitter to pixel, $x_2$ the sound path from pixel to receiver, and $c$ the sound velocity. In the TFM applied in this paper, only direct sound paths and p-waves are considered.

\subsection{Reverse Time Migration (RTM)}
\label{RTM}

Elastic RTM is a simulation-based imaging method, which solves the elastic wave equation forward in time. 
Subsequently, the difference between simulated and observed data is injected back into the domain and propagated backward in time, utilizing the reciprocity of the wave equation~\cite{KOCUR201696}. 
RTM identifies the positions and shapes of unknown scatterers (i.e. `defects') by convolving the forward and backward wave fields. For the position vector $\tensor{x} = \left[x, z\right]$ and solution vector $\mathbf{u} = \left[ u_\text{x}, u_\text{z} \right] \in \mathbb{R}^2$, the 2-dimensional elastic wave equation in the domain $\Omega \subset \mathbb{R}^2$ for time $T = [0, T_\text{e}]$ is
\begin{equation}
    \rho(\tensor{x})\ddot{\tensor{u}}(\tensor{x}, t) - \nabla \cdot \left(\tensor{C}(\tensor{x}) \colon \nabla \tensor{u}(\tensor{x}, t)\right) = \tensor{f}(\tensor{x}, t)\text{,} \quad \tensor{x} \in \Omega \ \text{,} \quad t \in T \label{eq:elasticwe}
\end{equation}
where $\rho$ represents the specimen's density, $\tensor{C}$ the linear isotropic plane strain constitutive tensor, and $\tensor{f}$ the external force vector. The initial conditions are set to $\tensor{u}(\tensor{x}, 0) = \dot{\tensor{u}}(\tensor{x}, 0) = \tensor{0}$. Additionally, boundary conditions are defined on $\partial \Omega$. Parameterized by the density~$\rho$, the p-wave speed $v_\text{p}$, and the s-wave speed $v_\text{s}$, the components of $\tensor{C}$ in an isotropic medium are expressed as
\begin{equation}
    C_{ijkl} = \rho(v_\text{p}^2 - 2 v_\text{s}^2) \delta_{ij} \delta_{kl} + \rho v_\text{s}^2 \delta_{ik} \delta_{jl} + \rho v_\text{s}^2 \delta_{il} \delta_{jk} \text{.}
\end{equation}

In the backward simulation, the source term $\tensor{f}^\dagger$ is defined as the difference between the simulated data $\tensor{u}(\tensor{x}, t)$ and observed data $\tensor{u}^\text{o}(\tensor{x}, t)$. 
The difference is evaluated and injected at the receiver positions $\tensor{x}^r$, $r=1, ..., N$,
\begin{equation}
    \tensor{f}^\dagger(\tensor{x}, t) = \sum_{r=1}^{N} \left( \tensor{u}(\tensor{x}, t) - \tensor{u}^\text{o}(\tensor{x}^r, t) \right)\delta(\tensor{x} - \tensor{x}^r) \cdot \tensor{n}^{r} \tensor{n}^{r}
\end{equation}
where $\tensor{n}^{r}$ is a unit vector pointing in the direction in which the sensor measures the wave field. Since the wave equation is self-adjoint, the backward wave field solves 
\begin{equation}
    \rho(\tensor{x})\ddot{\tensor{u}}^\dagger(\tensor{x}, t) - \nabla \cdot \left(\tensor{C}(\tensor{x}) \colon \nabla \tensor{u}^\dagger(\tensor{x}, t)\right) = \tensor{f}^\dagger(\tensor{x}, t)\text{,} \quad \tensor{x} \in \Omega\text{,} \quad t \in T
    \label{adjoint}
\end{equation}
with end conditions $\tensor{u}(\tensor{x}, T_\text{e}) = \dot{\tensor{u}}(\tensor{x}, T_\text{e})=\tensor{0}$. The backward wave field can be computed by starting at end time $T_e$ and integrating the solution backward in time. 
In the presented study, all wave simulations are solved with the spectral element method, a variant of the finite element method for time-dependent problems, in particular wave propagation problems.
Therefore, the solution vector is approximated by a set of basis functions.
After the  elastic wave equation is discretized in space, the semi-discrete system is integrated in time.
More information on the simulation setup are give in Section~\ref{simsetup}.

In elastic RTM~\cite{Rao2}, the zero-lag correlation of the forward wave field with the backward wave field is examined as follows
\begin{equation}
    \mathrm{RTM}(x, z) = \int_T \tensor{u}(\tensor{x}, t) \cdot \tensor{u}^\dagger(\tensor{x}, t) dt \text{.}
\end{equation}
However, in the study at hand, the RTM intensity is considered as the sensitivity kernel with respect to a density perturbation, due to its association with full waveform inversion. At the same time, the p- and s-wave speeds remain undisturbed. The connection to the first gradient of the corresponding FWI optimization problem is shown in the next subsection. The modified RTM image used in the following is introduced as
\begin{equation}
    \begin{split} &RTM_\rho(x, z) = \\ 
    &\int_T - \dot{\tensor{u}}(\tensor{x}, t) \cdot \dot{\tensor{u}}^\dagger(\tensor{x}, t) + \nabla \tensor{u} (\tensor{x}, t)\colon \tilde{\tensor{C}}(\tensor{x}, t) \colon \nabla \tensor{u}^\dagger(\tensor{x}, t) dt \end{split}
\end{equation}
where $\tilde{\tensor{C}}$ is the constitutive tensor derived with respect to the density
\begin{equation}
    \tilde{C}_{ijkl} = (v_\text{p}^2 - 2 v_\text{s}^2) \delta_{ij} \delta_{kl} + v_\text{s}^2 \delta_{ik} \delta_{jl} + v_\text{s}^2 \delta_{il} \delta_{jk} \text{.}
\end{equation}
For multiple experiments, the total RTM sums up the RTM values of the individual experiments. In the paper at hand, we evaluate the absolute value of the modified RTM on a grid and smooth it afterwards using a Gaussian filter with a standard deviation of 3. 

\subsection{Full Waveform Inversion (FWI)}
\label{FWI}

In contrast to RTM, FWI is an iterative data fitting method that updates the material parameters of a numerical model to match simulated wave fields to observed wave fields. 
The results in~\cite{Buerchner2023, BKK23, Rabinovich2024} show that inverting for the density while leaving the wave speeds of the material untouched is beneficial for the identification of high-contrast scatterers such as voids. 
This approach is adopted in the presented study. 
The underlying optimization problem is 
\begin{equation}
    \rho^*(\tensor{x}) = \arg \min_{\rho(\tensor{x})} \chi \left(\rho(\tensor{x})\right)
\end{equation}
where the cost function $\chi$ sums up the $L_2$ waveform misfit at all receivers 
\begin{multline}
    \chi \left(\rho(\tensor{x})\right) = \\ \frac{1}{2} \sum_{r=1}^{N} \int_T \int_\Omega \left( \left( \tensor{u}(\rho; \tensor{x}, t) - \tensor{u}^\text{o}(\tensor{x}^r, t) \right) \cdot \tensor{n}^{r} \right)^2 \delta\left(\tensor{x} - \tensor{x}^r\right) d\Omega dt \text{.}
\end{multline}
Again, $\tensor{u}$ denotes the forward wavefield solving equation~\eqref{eq:elasticwe} and $\tensor{u}^\text{o}$ is the measured wave field, while $\tensor{n}^r$ points in the measuring direction of the receivers. For multiple experiments, the individual misfits are summed up. By applying the adjoint method, the derivative of the cost function with respect to a perturbation of the density $\rho$ in direction $\delta \rho$ can be expressed as
\begin{equation}
    \nabla_\rho \chi \delta \rho = \int_\Omega K_\rho (\tensor{x}) \delta \rho d\Omega \text{.}
\end{equation}
The sensitivity kernel $K_\rho$ is equal to the just introduced modified RTM image
\begin{equation}
    K_\rho(\tensor{x}) = \int_T - \dot{\tensor{u}} \cdot \dot{\tensor{u}}^\dagger + \nabla \tensor{u} \colon \tilde{\tensor{C}} \colon \nabla \tensor{u}^\dagger dt \text{.}
\end{equation}
and $\tensor{u}^\dagger$ is the solution of the adjoint wave equation introduced in~\eqref{adjoint}. 

As described in Section~\ref{simsetup}, the forward and backward wave fields are computed using the spectral element method. The wave fields are approximated by a linear combination of polynomial basis functions. Accordingly, the density $\rho$ is represented by a set of $N_\rho$ basis functions 
\begin{equation}
    \rho(\tensor{x}) = \sum_{i=1}^{N_\rho} N_i(\tensor{x}) \hat{\rho}_i
\end{equation}
where $N_i$ denotes the basis functions, and $\hat{\rho}_i$ are the corresponding coefficients. 
In the resulting discrete optimization problem, the coefficients $\hat{\rho}_i$ are the optimization parameters. 
The optimization problem is solved with the wave field modeling and inversion software Salvus~\cite{Salvus} from the Mondaic AG, using a trust-region L-BFGS algorithm. 
For the interested reader, Fichtner~\cite{Fichtner} provides a detailed explanation of full waveform modeling and inversion.

\subsection{Quantitative Comparison of Images}
\label{MET}
\begin{figure}[htbp!]
  \begin{center}
    \includegraphics[width=0.25\textwidth]{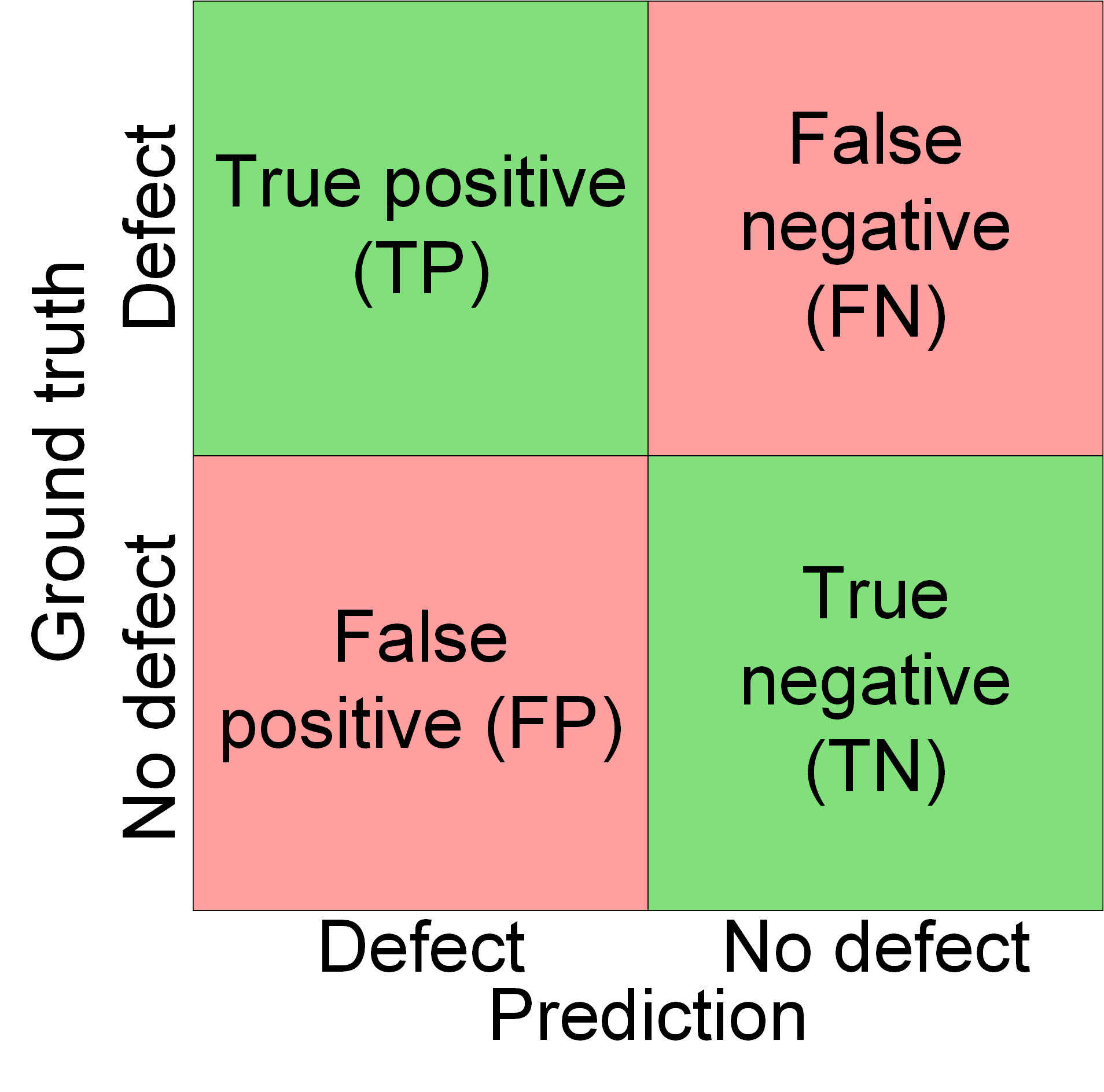}
  \end{center}
  \caption{Confusion matrix}
  \label{fig:confma}
\end{figure}

The study at hand compares the reconstruction approaches using different performance metrics derived by comparing a thresholded image to the ground truth. The four possible outcomes for each pixel are shown in the confusion matrix in Figure~\ref{fig:confma}. 

\begin{figure}[htb!]
    \centering
    \begin{subfigure}{0.23\textwidth}
        \centering
        \includegraphics[width=\textwidth]{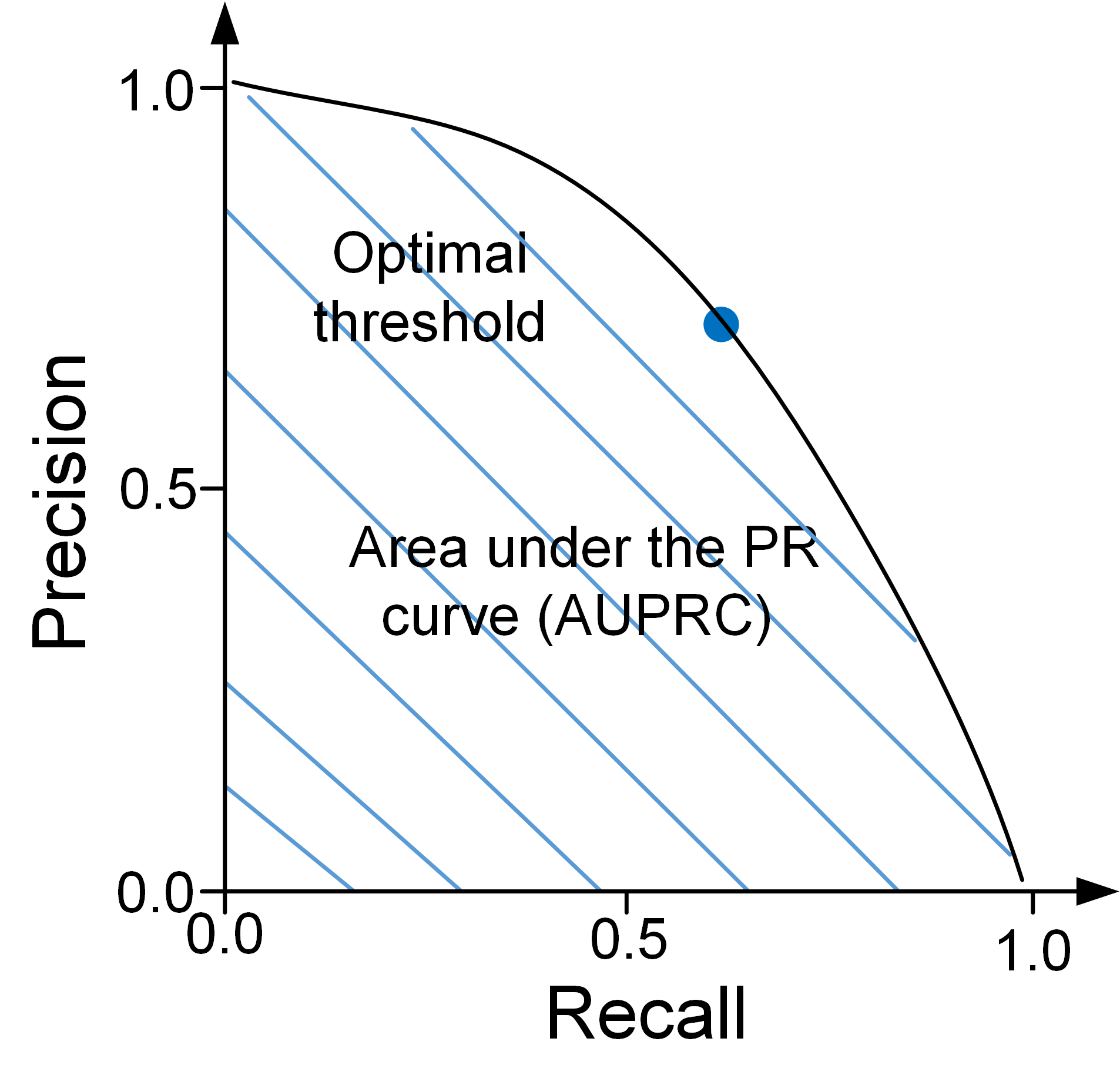}
        \caption{PRC}
    \end{subfigure}
    % \hfill
    \begin{subfigure}{0.23\textwidth}
        \centering
        \includegraphics[width=\textwidth]{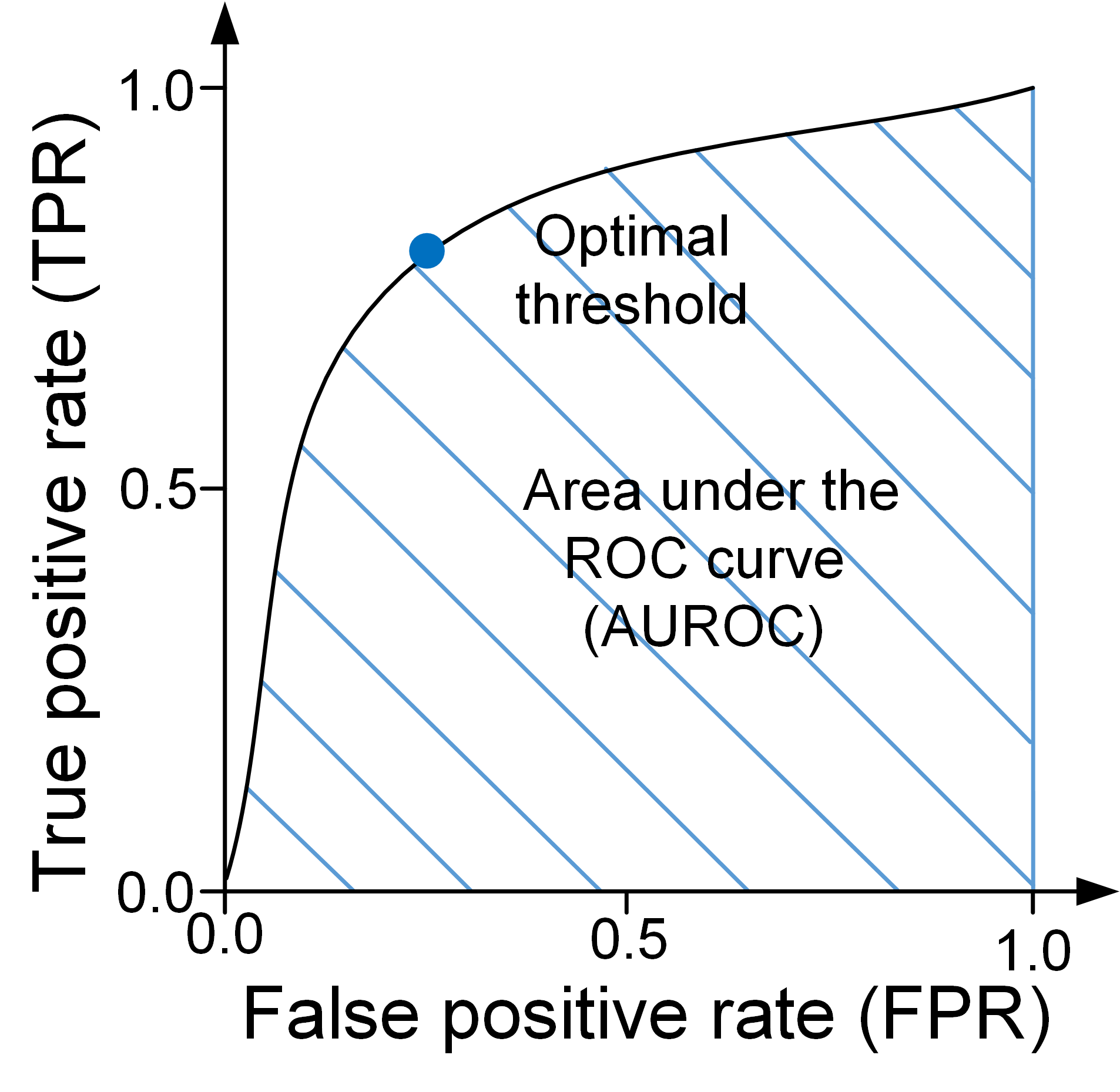}
        \caption{ROC}
    \end{subfigure}
	\caption{Schematic representation of the PRC and ROC according to~\cite{Jesse2006}}
    \label{fig:rocpr}
\end{figure}

Determining a suitable threshold for the images is a non-trivial task, as the threshold depends on how much importance is attributed to the four outcomes in the confusion matrix. In ultrasonic C-scans, the size of defects is commonly estimated by thresholding images using the $\SI{6}{\dB}$ criterion. However, this approach is not applicable to RTM or FWI since these methods visualize different quantities. A fixed threshold value for all three methods can, therefore, not be used. Instead, thresholds are determined for each image individually. The study at hand employs precision-recall curves (PRC) and receiver operating characteristic curves (ROC) to establish appropriate thresholds. Figure~\ref{fig:rocpr} provides sketches of exemplary PRC and ROC. 

The ROC plots the true positive rate (TPR) against the false positive rate (FPR) for varying threshold values. The TPR and FPR are defined as
\begin{equation}
\textrm{TPR} = \frac{\textrm{TP}}{\textrm{TP} + \textrm{FN}},
\end{equation}
and
\begin{equation}
\textrm{FPR} = \frac{\textrm{FP}}{\textrm{FP} + \textrm{TN}}.
\end{equation}
The PRC focuses on the ratio between precision and recall for different thresholds. Precision and recall are defined as
\begin{equation}
\textrm{precision}=\frac{\textrm{TP}}{\textrm{TP}+\textrm{FP}},
\end{equation}
and,
 \begin{equation}
\textrm{recall}=\frac{\textrm{TP}}{\textrm{TP}+\textrm{FN}}.
\end{equation}

With the ROC and PRC, thresholds can be determined automatically by identifying optimal points in the curves. For the ROC curve, this can be done by weighting TPR and FPR equally. The optimal threshold $\tau_{ROC}$ is defined as the one that minimizes the Euclidean distance from the point $(\textrm{FPR}, \textrm{TPR}) = (0, 1)$~\cite{bergstra2011}, which represents perfect classification
\begin{equation}
    \tau_\text{ROC} = \arg \min_\tau \sqrt{(1 - \textrm{TPR})^2 + \textrm{FPR}^2} \text{.}
\end{equation}
Similarly, for the PRC, perfect classification is reached at $(\textrm{precision}, \textrm{recall}) = (1, 1)$. Thus, the optimal threshold $\tau_{PRC}$ minimizes the Euclidean distance to that point
\begin{equation}
    \tau_\text{PRC} = \arg \min_\tau \sqrt{(1 - \textrm{precision})^2 + (1 - \textrm{recall})^2} \text{.}
\end{equation}

A further commonly used metric to estimate the performance in segmentation problems is the F1-score. The F1-score represents the harmonic mean of precision and recall, is particularly useful for imbalanced datasets, and is defined as
\begin{equation}
\textrm{F1-score}=2 \cdot \frac{\textrm{precision} \cdot \textrm{recall}}{\textrm{precision}+\textrm{recall}}= \frac{\textrm{TP}}{\textrm{TP}+0.5  \cdot (\textrm{FP}+\textrm{FN})}.
\end{equation}
In this study, the F1-score is further used to determine a threshold that maximizes its value. This process involves adjusting the threshold for classifying instances and selecting the threshold that yields the highest F1-score. This approach is compared against those based on the PRC and ROC in the qualitative comparison in Subsection~\ref{QUAL}. For more details on evaluating the performance of classification tasks, please refer to \cite{hastie2009, alpaydin2020}.

The maximal F1-score, the area under the ROC (AUROC), and the area under the PRC (AUPRC) are evaluated for a quantitative comparison. The AUROC and AUPRC take different thresholds into account and, therefore, provide information about the contrasts in the images and the robustness of classification.

\section{Experimental, Simulation, and Optimization Setup}
\label{SETUP}
\subsection{Investigated Specimens and Experimental Setup}
\label{expsetup}
\begin{figure*}[htbp!]
	\centering
	\includegraphics[trim=0cm 0cm 0cm 0cm,width=0.9\textwidth]{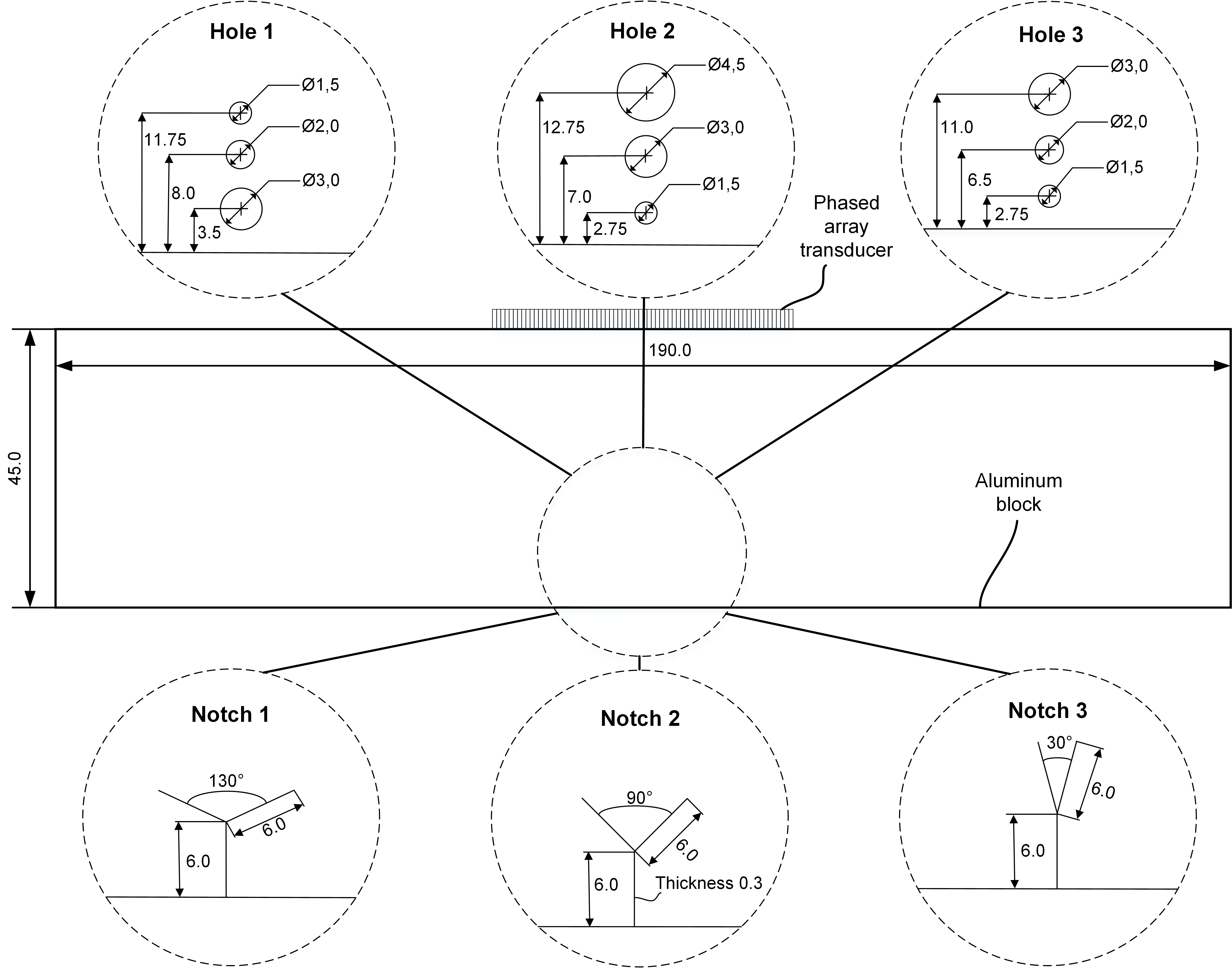} 
	\caption{Phased array setup with six different investigated defects; from the top left to bottom right hole 1, hole 2, hole 3, notch 1, notch 2, and notch~3}
	\label{fig:defects}
\end{figure*}

Two different types of defects were investigated: side-drilled holes and Y-shaped notches. For each of the defect types, three variants of the defects were introduced in aluminum blocks measuring $\SI{190}{\milli \meter} \times \SI{45}{\milli \meter} \times \SI{50}{\milli \meter}$. The side-drilled holes were manufactured with conventional drilling, while the Y-shaped notches were produced using electrical discharge machining. The different defects are illustrated in Figure~\ref{fig:defects}. The design of the Y-shaped notches is inspired by the study of Rao et al.~\cite{Rao1}.

The aluminum specimens consist of the alloy AlMg3. The ultrasound p- and s-wave speeds, $v_{\textrm{p}}$ and $v_{\textrm{s}}$, were determined using the OmniScan MX2 ultrasonic device from the company Olympus (Evident). The measurements utilized a $\SI{2.25}{\mega \hertz}$ p-wave transducer (V104-RM) and a $\SI{5}{\mega \hertz}$ s-wave transducer (V157-RM), both from Olympus (Evident). Multiple echoes were used to determine the wave speeds, allowing the measurement to be independent of the latency of the measurement system. The estimated speeds were $v_{\textrm{p}} = \SI{6315.8}{\meter \per \second}$ and $v_{\textrm{s}} = \SI{3129.3}{\meter \per \second}$. Additionally, the density of the specimens was calculated, by weighting the specimens and taking their nominal dimensions, resulting in $\rho = \SI{2582.8}{\kilogram \per \meter^3}$. 

The AOS OEM-PA ultrasonic device was used as an acquisition system for the FMC dataset. It has 64 channels with a 12-bit digitization. As a transducer, the Olympus 2.25L64-A2 probe was used. It has a central frequency of $\SI{2.25}{\mega \hertz}$, a bandwidth of 75\%, and 64 elements with a pitch of $\SI{0.75}{\milli \meter}$. The aperture's width is $\SI{47.25}{\milli \meter}$. The acquisition software was ARIA from the Phased Array Company (TPAC).

\subsection{Simulation Setup}
\label{simsetup}
As previously mentioned, all wave simulations in this study are conducted using Salvus~\cite{Salvus} from the Mondaic AG. 
The wave simulation kernel employs the spectral element method~(SEM)~\cite{Komatitsch1999}, a variant of the finite element method~(FEM)~\cite{Hughes2012}. 
Lagrange polynomials that interpolate Gauss-Lobatto-Legendre (GLL) points serve as shape functions to approximate the wavefield. 
SEM achieves a diagonal mass matrix using GLL points also as quadrature points to integrate the weak form. The resulting semi-discrete system is integrated explicitly in time using second-order central differences~(CDM). All simulations neglect damping and solve the 2D elastic wave equation to approximate the experimental setup depicted in Figure~\ref{fig:defects}.
% omit damping effects and use a 2D plain strain model to approximate the experimental setup depicted in Figure~\ref{fig:defects}.
The synthetic reference simulations use boundary-conforming meshes that adopt the shapes of the distinct defects. To avoid inverse crime, the spatial and temporal resolution of the reference simulations is considerably higher than the resolution in the optimization simulations. The piezoelectric transmitters are modeled as point sources, exciting the specimen perpendicularly to its surface. In the RTM and FWI simulations, the numerical model incorporates absorbing boundaries on both sides to reduce the overall computation time. The modeled physical domain measures $\SI{67.25}{\milli \meter} \times \SI{45}{\milli \meter}$.

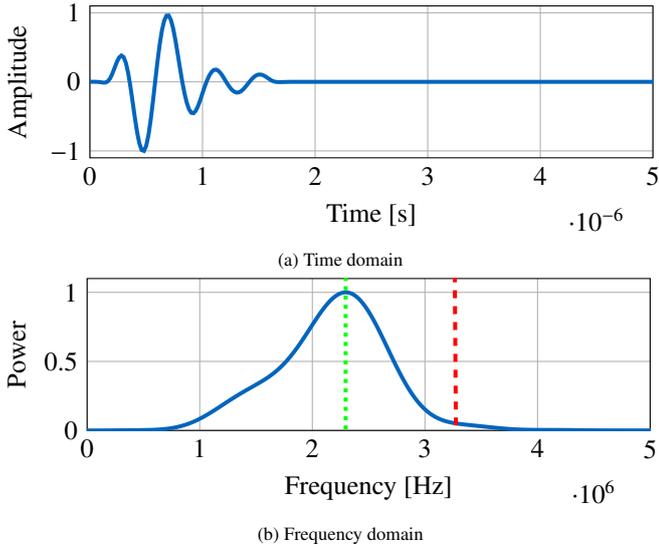
\begin{figure}[htb!]
    \begin{subfigure}{0.49\textwidth}
        \definecolor{TumBlue}{RGB}{0,101,189} % blue
\definecolor{Orange}{RGB}{227,114,34} % orange
\definecolor{lightgray204}{RGB}{204,204,204}
\definecolor{darkgray176}{RGB}{176,176,176}
\begin{tikzpicture}
	\begin{axis}[
		xmin = 0.0e-6, xmax = 5e-6,
		ymin = -1.1, ymax = 1.1,
        xtick style={color=black},
		ytick={-1, 0, 1},
        x grid style={darkgray176},
        y grid style={darkgray176},
        xmajorgrids,
        ymajorgrids,
        xtick style={draw=none},
        ytick style={draw=none},
		width = \textwidth,
		height = 0.4\textwidth,
		xlabel = {Time $[\si{\second}]$},
		ylabel style={align=center}, 
		ylabel={Amplitude}
		]
		
		\addplot[TumBlue, line width=1.5pt] file[] {Figures/stf_t.dat};
	\end{axis}
\end{tikzpicture}%
        \caption{\centering Time domain}
    \end{subfigure}
    \begin{subfigure}{0.49\textwidth}
        \definecolor{TumBlue}{RGB}{0,101,189} % blue
\definecolor{Orange}{RGB}{227,114,34} % orange
\definecolor{lightgray204}{RGB}{204,204,204}
\definecolor{darkgray176}{RGB}{176,176,176}
\begin{tikzpicture}
	\begin{axis}[
		xmin = 0.0e6, xmax = 5e6,
		ymin = 0.0, ymax = 1.1,
        xtick style={color=black},
		ytick={0, 0.5, 1},
        x grid style={darkgray176},
        y grid style={darkgray176},
        xmajorgrids,
        ymajorgrids,
        xtick style={draw=none},
        ytick style={draw=none},
		width = \textwidth,
		height = 0.4\textwidth,
		xlabel = {Frequency $[\si{\hertz}]$},
		ylabel style={align=center}, 
		ylabel={Power},
        ]
        
		\addplot[TumBlue, line width=1.5pt] file[] {Figures/stf_f.dat};
  
      \addplot[line width=1.5pt, green, dotted] coordinates {(2.296e6, -1) (2.296e6, 3)};
  
      \addplot[line width=1.5pt, red, dashed] coordinates {(3.284e6, -1) (3.248e6, 3)};
	\end{axis}
\end{tikzpicture}%
        \caption{\centering Frequency domain}
    \end{subfigure}
	\caption{Normalized source time function in time and frequency domain; green dotted line marks maximal power and red dashed line marks power below which $\SI{95}{\percent}$ of the signal's energy is contained}
    \label{stf}
\end{figure}

For the synthetic inversions, the density and wave speeds of the reference and optimization models were set according to the measurements detailed in the previous subsection. During the inversions on the experimental data, a slight mismatch in the phase of the first and second reflected p-waves was observed in the simulations. To address this discrepancy, the p-wave speed of the optimization model was adjusted to be $v_\text{p} = \SI{6360}{\meter \per \second}$ in the corresponding RTMs and FWIs on experimental data. The excitation function of the piezoelectric transducers was estimated by isolating the first reflection from a measurement signal and applying a time filter to it. Figure~\ref{stf} illustrates the source time signal alongside its normalized frequency content.
The frequency with maximum power in the excitation signal is $f_\text{max} = \SI{2.296}{\mega \hertz}$. The frequency below which $\SI{95}{\percent}$ of the signal's energy is contained is $f_\text{95} = \SI{3.284}{\mega \hertz}$.
The inversion model is discretized to accurately resolve wavelengths of the s-waves up to $f_\text{95}$ with $1.5$ quartic elements per wavelength, resulting in a mesh with $9\,869$ elements and $158\,745$ nodes. The simulations are performed until $T_\text{e} = \SI{2.5e-5}{\second}$ with a fixed time step size of $\Delta t = \SI{1.0e-9}{\second}$, as the critical time step size may decrease drastically for density-scaling during the FWI process.

\subsection{Optimization Setup}

For the FWI, a region of interest~(ROI) is defined, with the density constrained to be between $0.1$ and $1.0$ times the background density. The lower bound is introduced to ensure stability in the simulations. The specifications are listed in Table~\ref{tab:fwi}. 
\begin{table}[htbp!]
\centering
\caption{FWI optimization specifications}
\resizebox{0.3\textwidth}{!}{
\begin{tabular}[t]{cc}
ROI $x$ & $[\SI{70}{\milli \meter}, \SI{120}{\milli \meter}]$\\
\hline
ROI $y$ & $[\SI{0}{\milli \meter}, \SI{25}{\milli \meter}]$\\
\hline
bounds $\rho$ & $[\SI{258.28}{\kilogram \per \meter^3}, \SI{2582.8}{\kilogram \per \meter^3}]$
\label{tab:fwi}
\end{tabular}
}
\end{table}
The entire FWI process is structured into two distinct optimization runs. The first run comprises 20 FWI iterations. Source stacking is employed to reduce computational costs, which involves superimposing the data of eight individual experiments based on the positions of the phased array sources. Therefore, the number of forward simulations required to compute the gradient is reduced to eight. An additional eight adjoint simulations have to be conducted to compute the gradient. Data containing the first p-wave reflections from the back wall are excluded in this first run. 

The second run of the FWI process builds on the previous results, by utilizing the first run's final model as its initial model. 
To improve the reconstruction quality, any back wall artifacts of this model are removed, and all densities above $0.9$ times the original value are reset to background density. 
The second run concludes either when no better solution is found or after a maximum of 20 additional iterations. 
In the second run, the stacking of sources is reduced to two sources per supershot, and the entire wave signals are utilized. 
The study's outcome demonstrates that this two-step approach significantly enhances the reconstructing quality. 
The first run rather identifies the vertical shape of the defects, while the second run further improves the estimation of the defect's width, by incorporating the back wall reflection data. 

\section{Results and Discussion}
\label{RESULTS}
\subsection{Qualitative Comparison of the Methods}
\label{QUAL}

\begin{figure*}[htbp!]
	\centering
	\includegraphics[trim=0cm 0cm 0cm 0cm,width=0.95\textwidth]{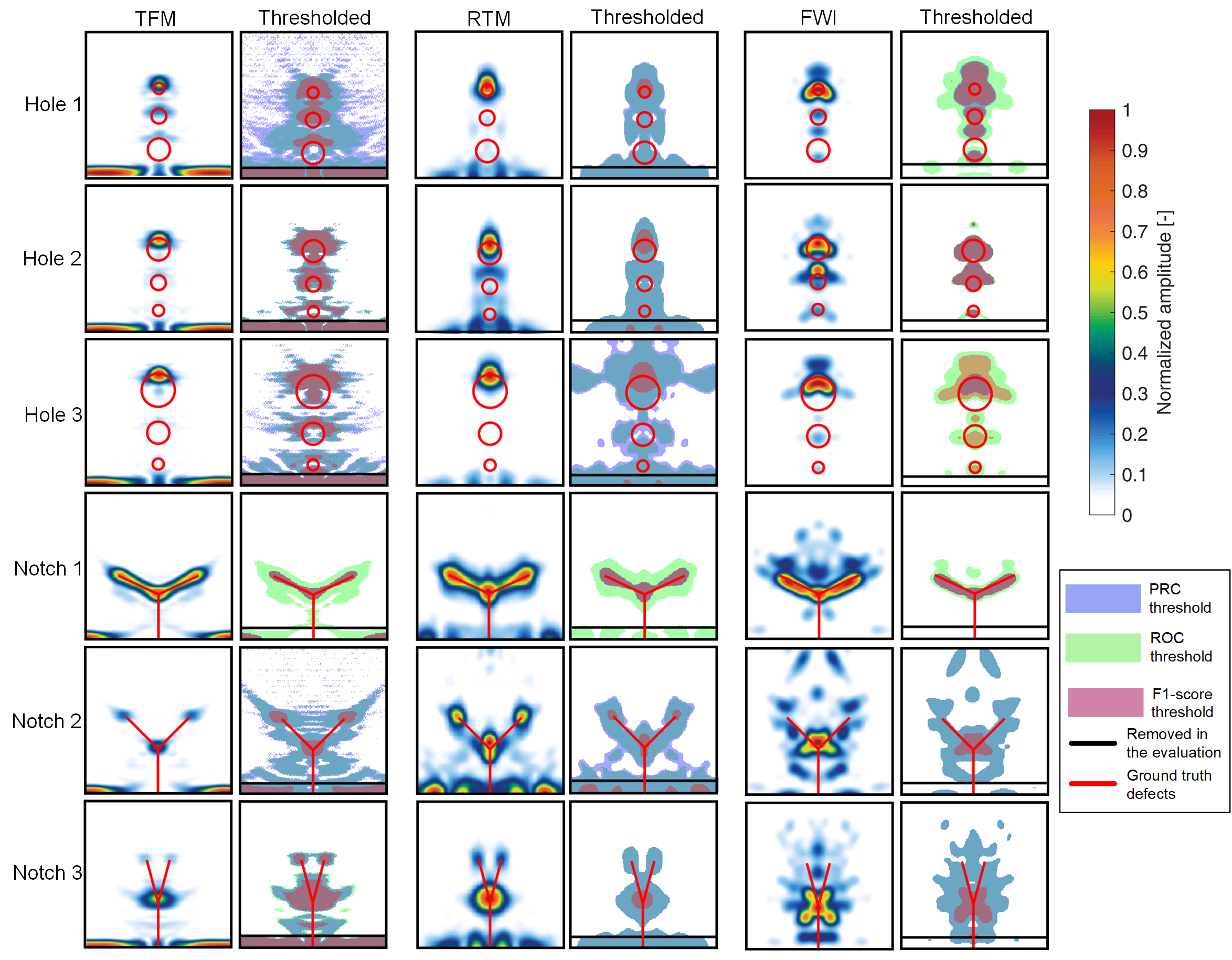} 
	\caption{Reconstructed images of the synthetic dataset: next to the reconstructed images the thresholded images based on the PCR curve, ROC curve, and F1-score are shown; defect boundaries are highlighted in red}
	\label{fig:ressyn}
\end{figure*}

\begin{figure*}[htbp!]
	\centering
	\includegraphics[trim=0cm 0cm 0cm 0cm,width=0.95\textwidth]{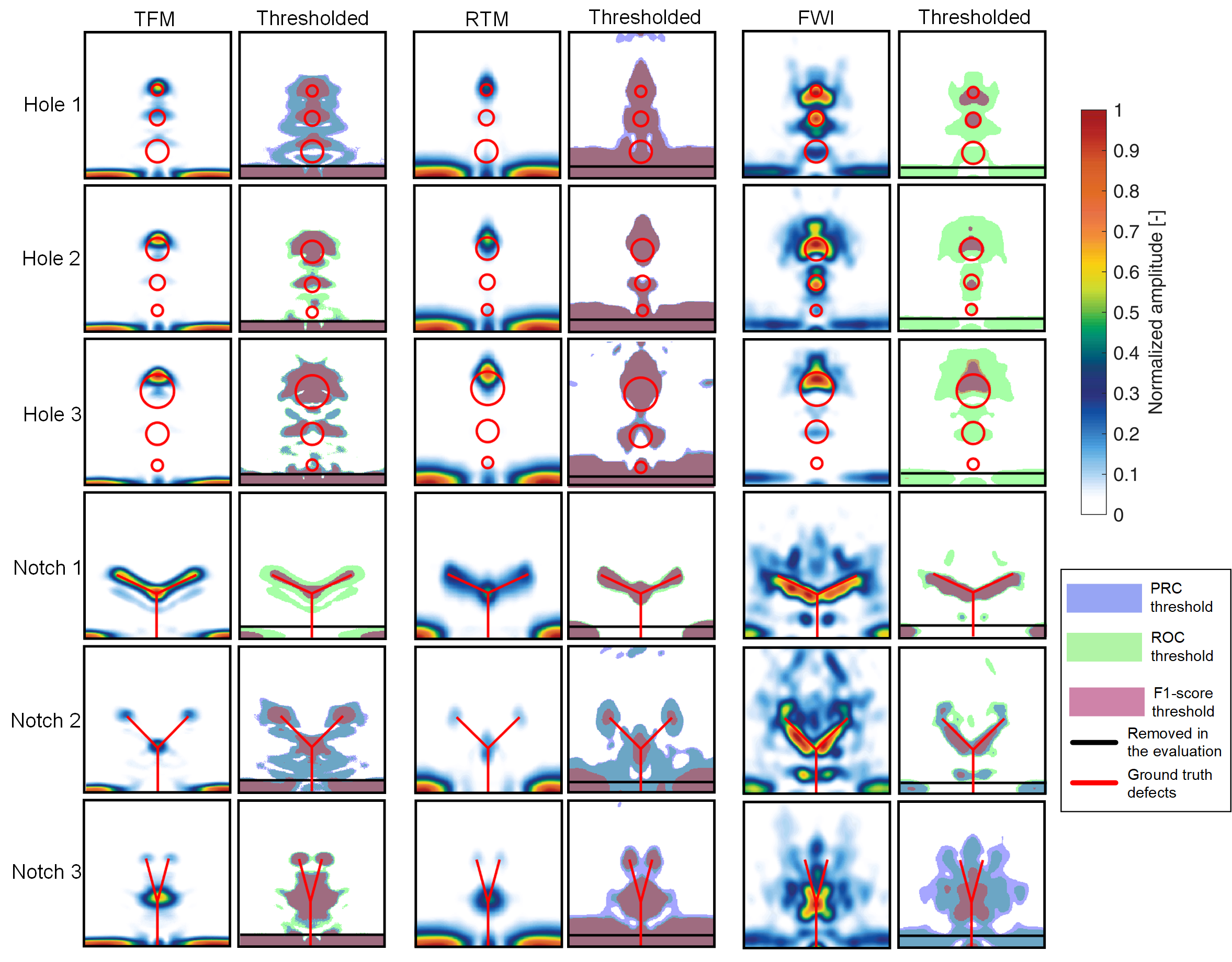} 
	\caption{Reconstructed images of the experimental dataset: next to the reconstructed images the thresholded images based on the PCR curve, ROC curve, and F1-score are shown; defect boundaries are highlighted in red}
	\label{fig:resexp}
\end{figure*}

The synthetic and experimental datasets were reconstructed using the three different methods. The resulting images are displayed in Figures~\ref{fig:ressyn} and~\ref{fig:resexp}. 

The boundaries of the ground truth for the defects are highlighted in red. The thresholded images are shown in the right columns next to the reconstructed images in the left columns. The thresholds were determined based on the PRC, ROC, and the maximal F1-score. The thresholded images are plotted transparently over one another. When determining the thresholds, the bottom of the images, indicated by black lines and corresponding to back wall echos, was not considered. This was done because the back wall echoes vary with the applied reconstruction methods and often have large amplitudes, which distort the evaluation. 

The F1-score typically gives the highest threshold and, therefore, is the least sensitive. For the PRC and ROC, in some cases, one or the other threshold is higher. Especially, the PRC gives a comparatively lower threshold in some cases, estimating most of the image as a defect (e.g., for the synthetic dataset in Figure~\ref{fig:ressyn}, hole 1, TFM).

In the reconstructed non-thresholded images of the holes, it is evident that the upper defects overshadow the bottom ones, resulting in less information about the lower defects in the reconstructed images. 
This issue is more pronounced when the top defect is larger than the bottom, as seen in hole 3. Nevertheless, the bottom defects are most clearly visible in the FWI reconstruction. Additionally, the position and shape of the defects are better captured by FWI compared to TFM and RTM. In TFM and RTM, the defect indications tend to shift upwards.

For notch 1, all methods show a qualitatively good result and the V-shape is clearly visible with all reconstruction methods. However, notches 2 and 3 were more challenging to reconstruct for all methods. For these notches, the ultrasonic energy tends to be reflected away from the phased array transducer and mode conversion lead to later arriving wave modes, which are not considered in the direct TFM reconstruction. In the TFM and RTM images, the diffraction at the top and bottom of the notch is mostly visible. An exception is the reconstruction of notch 2 from the experimental dataset with FWI, where the V-shape was successfully captured. This can be explained by the fact that FWI utilizes more information within the signal than just the diffraction echo. RTM is primarily sensitive to the dominant diffraction echo and does not capture subtle changes in the signal that the iterative FWI approach can detect. The TFM and RTM reconstruction of notch 3 show a pronounced indication on the bottom of the V-shape, but still the diffraction tips on top. The FWI reconstruction of notch 3 also show a pronounced indication on the bottom part, but the V-Shape is not recognizable. 
The vertical notch below the V-shape could not be resolved by any of the methods as a result of the aperture of the phased array sensor and placing it directly above the defects.

The thresholded images in general are very sensitive to the defect type and reconstruction algorithm. F1-score shows good results for TFM images, but overestimates defects in RTM. It also seems that F1-score is less sensitive to small defects as they are not segmented properly.

In contrast to one's intuition, FWI reveals more details of the defects in the experimental datasets compared to the synthetic datasets. Although the optimization workflow used for both datasets was the same, the optimization process is trapped earlier in a local minimum for the synthetic data. This observation highlights the importance of properly incorporating prior knowledge and regularization, as the FWI optimization problem underlying this phased array testing measurement setup is inherently ill-posed. 

\subsection{Quantitative Comparison of the Methods}
\label{COM}

Three performance metrics were used to quantitatively assess the reconstruction quality, as outlined in Section~\ref{MET}. The calculated metrics include the maximal F1-score derived from the thresholded images as shown in Subsection~\ref{QUAL}. The thresholds based on PRC and ROC were not further considered, as they tended to be overly sensitive. Instead, AUPRC and AUROC metrics were used, as they account for various thresholds and therefore better reflect the contrast in the image. When calculating the metrics, the area close to the back wall was removed. The metrics for both the synthetic and experimental datasets are presented in Tables~\ref{tab:metsynth} and~\ref{tab:metexp}. The best performing reconstruction approach for each metric was marked in bold.

\begin{table}[htbp!]
\centering
\caption{Metrics on the synthetic dataset.}
\resizebox{\linewidth}{!}{
\begin{tabular}[t]{lcccccc}
\hline
Metric/Specimen&Hole 1& Hole 2 & Hole 3 & Notch 1 & Notch 2 & Notch 3\\
\hline
F1 TFM& 0.146& 0.223& 0.271& \textbf{0.587}& 0.189&0.228\\
F1 RTM& 0.223& 0.345& 0.320& 0.549& 0.238&\textbf{0.287}\\
F1 FWI&\textbf{0.305} &\textbf{0.567} & \textbf{0.485}& 0.581& \textbf{0.252}&0.254\\
AUPRC TFM& 0.034& 0.084& 0.086& 0.341& 0.062 &0.081\\
AUPRC RTM& 0.145& 0.271& 0.203& \textbf{0.487}&0.105 & \textbf{0.271}\\
AUPRC FWI& \textbf{0.232}& \textbf{0.529}& \textbf{0.490}& 0.443& \textbf{0.147} &0.176\\
AUROC TFM& 0.730 & 0.860&0.758 &0.922 &0.786 &0.867 \\
AUROC RTM&0.862& 0.911& 0.789& \textbf{0.946}& \textbf{0.868}& \textbf{0.918}\\
AUROC FWI&\textbf{0.900} & \textbf{0.978}& \textbf{0.923}&0.853 & 0.834&0.867\\
\hline
\label{tab:metsynth}
\end{tabular}
}
\end{table}

\begin{table}[htbp!]
\centering
\caption{Metrics on the experimental dataset.}
\resizebox{\linewidth}{!}{
\begin{tabular}[t]{lcccccc}
\hline
Metric/Specimen&Hole 1& Hole 2 & Hole 3 & Notch 1 & Notch 2 & Notch 3\\
\hline
F1 TFM&0.192 &0.264 & 0.308& \textbf{0.566}& 0.206&0.232\\
F1 RTM&0.170 & 0.210& 0.267&0.374 &0.141 &0.168\\
F1 FWI& \textbf{0.431}& \textbf{0.580}&\textbf{0.452} & 0.391&\textbf{0.351} &\textbf{0.313}\\
AUPRC TFM& 0.064& 0.094&0.126 &\textbf{0.314} &0.071 &0.083\\
AUPRC RTM& 0.040& 0.062& 0.075& 0.138& 0.038&0.049\\
AUPRC FWI& \textbf{0.332}& \textbf{0.602}& \textbf{0.404}& 0.172& \textbf{0.192}&\textbf{0.211}\\
AUROC TFM& 0.829& 0.878& 0.820& \textbf{0.933}& 0.830& \textbf{0.875}\\
AUROC RTM& 0.806& 0.854&0.785 & 0.901& 0.790&0.827\\
AUROC FWI& \textbf{0.915}&\textbf{0.939} & \textbf{0.873}& 0.810& \textbf{0.853}&0.854\\
\hline
\label{tab:metexp}
\end{tabular}
}
\end{table}

For the holes, all performance metrics indicate that FWI yields the most accurate reconstructions for both the synthetic and experimental datasets. Nevertheless, the thresholded images by PRC and F1 do not segment all 3 defects for hole configuration 1, 2 and 3. ROC threshold overestimates the side drilled holes in all 3 configurations. In the case of RTM the F1-score threshold can segment the side drilled holes in all three configurations. In the experimental dataset concerning notch 2, FWI also outperforms the other methods in all metrics. However, for notch 1, TFM shows the highest metrics. For the synthetic dataset, the trend between the metrics and the reconstruction methods is less clear for the notches, but FWI also reaches the highest metric in the F1-score for notch 2. The higher metrics for TFM and RTM on the synthetic dataset are due to their better reconstruction of the tips and the bottom of the V-shape.

In case of the notches in the experimental dataset, FWI shows for notch 2 the highest metrics. For notch 1 TFM yields best results for all metrics. The situation is not so clear for notch 3, TFM yields highest AUROC, but FWI best F1 and AUPRC metric. Notch 3 is the most challenging testing case, because of the small angle of the V-shape. Here, TFM and RTM only detects the tips and FWI detects only the bottom of the V-shape.

In general, it should be noted that in reconstructed ultrasonic images, only the top boundary of a larger defect is captured, while the inner region of the defect cannot be resolved. Therefore, the evaluation metrics primarily indicate how accurately this top boundary is resolved.

%- Metriken zeigen, dass für verdeckte Bohrungen, FWI die besten ergebnisse erzielt.

%F1-score threshold zeigt bei den Bohrungen in RTM Probleme Reflektoren zu identifitzieren.

%PRC Therhohlding zeigt bei Bohrungen in TFM eher noise an.

%Wo ist ROC bei TFM/ RTM for Hole 1 , 2, 4? Nicht sichtbar.

%ROC threshold bei FWI Bohrungen: dreimal so groß wie eigentlicher Defekt?

\section{Conclusions}
\label{CON}

In this work, we investigate three different imaging or reconstruction algorithms, namely TFM, RTM, and FWI, for phased array FMC data, and compare their performance qualitatively and quantitatively. 
Hereby, the localization and size determination of side-drilled holes and notches in an aluminum specimen are assessed.
The comparison includes both visual assessments and segmentation metrics.
While TFM and RTM show similar performance, FWI better captures the positions of most defects, provides higher contrast in the images (as indicated by AUROC and AUPRC), and can reveal some aspects of defects where little information is available due to shadowing by upper defects. 
However, FWI also encounters difficulties in scenarios where multiple reflections, diffraction, and mode conversions occur. 
The V-shaped parts of the notches cause the superposition of various wave modes, leading to inaccurate reconstructions and imaging artifacts.
Nevertheless, for one notch (notch 2 of the experimental dataset), FWI demonstrates superior performance compared to the other two methods, which shows that even the 45 degree legs of the notch can be fully imaged.
Whereas in TFM an RTM reconstruction, only the diffraction signals are visible.
However, the improved reconstruction performance of FWI comes at the cost of significantly increased computational complexity. 

The results highlight the potential of FWI for crack sizing in diffraction-driven ultrasonic testing methods.
The superior performance of RTM compared to TFM and SAFT, as demonstrated in the literature~\cite{Rao2, Grohmann}, could not be confirmed for the investigated defect configurations in this study. 
In general, it can be concluded that the specific defect type, together with the measurement configuration (e.g., aperture size), strongly affects the reconstruction quality.
Furthermore, it should be noted that these defects represent artificial defects, and real defects could have arbitrary shapes. 
In addition, the material is homogeneous, and investigating a heterogeneous material would add another level of complexity.

Finally, it is surprising that FWI provides better overall reconstructions for the experimental datasets than for the synthetic datasets.
The optimization problem is inherently ill-posed, causing it to converge to local minima.
This observation motivates future research to investigate the benefits of using additional regularization methods.
Further, the incorporation of neural networks for material discretization shows promising improvements in reconstruction quality~\cite{Herrmann2023}, but remains to be investigated for experimental datasets.
 
% For FWI, properly setting up the optimization problem is crucial to achieve accurate results. The better performance of FWI on the experimental dataset, compared to the synthetic dataset, highlights the importance of regularization for the underlying problem. Incorporating prior knowledge derived from the TFM image may also be advantageous. Additionally, mode separation into p- and s-waves may mitigate cross-talk~\cite{Rao2}.

%Furthermore, it is essential to accurately approximate the excitation for experimental measurements. An accurate source model can be found by inverting the source time function, as shown in Aktharuzzaman et al.~\cite{Aktharuzzaman2024}.

%=================================


\paragraph*{Acknowledgments}
The authors thank Dr. Lion Krischer and Dr. Christian Böhm from the Mondaic AG for their support. Furthermore, Tim Bürchner gratefully acknowledges funds received by the Deutsche Forschungsgemeinschaft under Grant no. KO 4570/1-2 and RA 624/29-2 (both grant number 438252876). We would also like to express our appreciation to Thomas Heckel and Christian Hassenstein for their assistance in conducting phased array measurements, although these were ultimately not used in the study.

\bibliographystyle{elsarticle-harv}
%\bibitem{Thurston}
%R. Thurston, A. Pierce,
%Ultrasonic Measurement Methods,
%Academic Press, London, England (1990)
%\bibitem{Kraut}
% J. Krautkramer, H. Krautkramer, Ultrasonic Testing of Materials, Springer, Berlin, Germany (1990)
\bibliography{main.bbl}





% \clearpage
% \tableofcontents 
% \end{thebibliography}
\end{document}